\documentclass[fleqn]{an}
\usepackage{graphicx}
\usepackage{times}
\usepackage{bm}
\Pagespan{73}{}
\Yearpublication{2010}
\Yearsubmission{2009}
\Month{9}
\Volume{331}
\Issue{1}
\DOI{10.1002/asna.200911252}
\begin{document}
\sloppy

\newcommand{\EQ}{\begin{equation}}
\newcommand{\EE}{\end{equation}}
\newcommand{\EQA}{\begin{eqnarray}}
\newcommand{\EEA}{\end{eqnarray}}
\newcommand{\brac}[1]{\langle #1 \rangle}
\newcommand{\pd}{\partial}
\newcommand{\pdz}{\partial_z}
\newcommand{\DIV}{\vec{\nabla} \cdot }
\newcommand{\CURL}{\vec{\nabla} \times }
\newcommand{\cross}[2]{\boldsymbol{#1} \times \boldsymbol{#2}}
\newcommand{\crossm}[2]{\brac{\boldsymbol{#1}} \times \brac{\boldsymbol{#2}}}
\newcommand{\ve}[1]{\boldsymbol{#1}}
\newcommand{\mean}[1]{\overline{#1}}
\newcommand{\meanv}[1]{\overline{\bm #1}}
\newcommand{\cst}{c_{\rm s}^2}
\newcommand{\nut}{\nu_{\rm t}}
\newcommand{\etat}{\eta_{\rm t}}
\newcommand{\etatz}{\eta_{\rm t0}}
\newcommand{\etaT}{\eta_{\rm T}}
\newcommand{\urms}{u_{\rm rms}}
\newcommand{\Urms}{U_{\rm rms}}
\newcommand{\brms}{B_{\rm rms}}
\newcommand{\Beq}{B_{\rm eq}}
\newcommand{\eu}{\hat{\bm e}}
\newcommand{\xu}{\hat{\bm x}}
\newcommand{\yu}{\hat{\bm y}}
\newcommand{\zu}{\hat{\bm z}}
\newcommand{\Ou}{\hat{\bm \Omega}}
\newcommand{\kef}{k_{\rm f}}
\newcommand{\tauc}{\tau_{\rm c}}
\newcommand{\tauto}{\tau_{\rm to}}
\newcommand{\St}{{\rm St}}
\newcommand{\Sh}{{\rm Sh}}
\newcommand{\Pm}{{\rm Pm}}
\newcommand{\Rm}{{\rm Rm}}
\newcommand{\Pra}{{\rm Pr}}
\newcommand{\Ra}{{\rm Ra}}
\newcommand{\Ro}{{\rm Ro}}
\newcommand{\Rey}{{\rm Re}}
\newcommand{\Co}{{\rm Co}}
\newcommand{\Cost}{\Omega_\star}
\newcommand{\ReLS}{{\rm Re}_{\rm LS}}
\newcommand{\qxx}{Q_{xx}}
\newcommand{\qyy}{Q_{yy}}
\newcommand{\qzz}{Q_{zz}}
\newcommand{\qxy}{Q_{xy}}
\newcommand{\qxz}{Q_{xz}}
\newcommand{\qyz}{Q_{yz}}
\newcommand{\qij}{Q_{ij}}
\newcommand{\Omx}{\Omega_x}
\newcommand{\Omz}{\Omega_z}
\newcommand{\emf}{\bm{\mathcal{E}}}
\newcommand{\emfi}{\mathcal{E}_i}
\newcommand{\nab}{\mbox{\boldmath $\nabla$} {}}
\newcommand{\meanFFFF}{\overline{\mbox{\boldmath ${\cal F}$}}{}}{}
\def\onethird{{\textstyle{1\over3}}}
\def\onehalf{{\textstyle{1\over2}}}
\def\threefourths{{\textstyle{3\over4}}}

\title{Convective dynamos in spherical wedge geometry}

\author{P.J. K\"apyl\"a\inst{1,2}\fnmsep\thanks{Corresponding author:
    {petri.kapyla@helsinki.fi}}\and M.J. Korpi\inst{1} \and
  A. Brandenburg\inst{2,3} \and D. Mitra\inst{4} \and R. Tavakol\inst{4}}

\titlerunning{Convective dynamos in spherical wedge geometry}
\authorrunning{P.J. K\"apyl\"a et al.}

\institute{ 
Observatory, University of Helsinki, PO BOX 14, FI-00014 University of 
Helsinki, Finland
\and
NORDITA, AlbaNova University Center, Roslagstullsbacken 23, SE-10691 
Stockholm, Sweden
\and
Department of Astronomy, AlbaNova University Center,
Stockholm University, SE-10691 Stockholm, Sweden
\and
Astronomy Unit, School of Mathematical Sciences, Queen Mary 
University of London, Mile End Road, London E1 4NS, United Kingdom}

\received{2009 Sep 9} 
\accepted{2009 Nov 16}
\publonline{2009 Dec 30}

\keywords{Sun: magnetic fields -- magnetohydrodynamics (MHD)}

\abstract{%
  Self-consistent convective dynamo simulations in wedge-shaped
  spherical shells are presented.
  Differential rotation is generated by the interaction of convection
  with rotation.
  Equatorward acceleration and dynamo action
  are obtained only for sufficiently rapid rotation.
  The angular velocity tends to be constant along cylinders.
  Oscillatory large-scale fields are found to migrate in the poleward
  direction.
  Comparison with earlier simulations in full spherical shells and
  Cartesian domains is made.}

\maketitle

\section{Introduction}
\label{sec:intro}
The large-scale magnetic fields in the Sun and other stars are thought
to arise from a complex interplay between nonuniform rotation,
stratification, and convective turbulence (e.g.\
Ossendrijver 2003; Brandenburg \& Subramanian 2005). According to
mean-field theory of turbulent dynamos, one of the main mechanisms
responsible for the generation of large-scale fields is the $\alpha$ effect,
which is related
to the kinetic helicity of the flow in idealised cases (Moffatt 1978;
Krause \& R\"adler 1980; R\"udiger \& Hollerbach 2004). In the
simplest settings, kinetic helicity is produced by rotating
inhomogeneous turbulence. The inhomogeneity
can arise from density stratification, or, in the absence of
stratification, from impenetrable boundaries. However, although
many early numerical simulations of rotating convection did work as 
dynamos
(e.g.\ Nordlund et al.\ 1992; Brandenburg et al.\ 1996), 
the magnetic field generated in these models was found to be 
mostly on small scales, with the typical spatial scale of the field
of the same order as the typical size of convection cells.
This posed a puzzle, given that the simulations included all the major
ingredients (stratification and rotation) known to be necessary
for the $\alpha$ effect to occur. 
Crucial insight into this problem was gained when the test-field method
(Schrinner et al.\ 2005, 2007) was used to compute the relevant
turbulent transport coefficients (K\"apyl\"a et al.\ 2009b). It was
found that the $\alpha$ effect resulting from the relatively slow 
rotation used in the earlier studies was not enough to overcome
turbulent diffusion. 
On the other hand, for more rapid rotation, turbulent diffusion was found to 
decrease which, according to mean-field models, would imply that 
large-scale dynamo action should become possible. 
This was recently confirmed by direct
simulations in the same parameter regime (K\"apyl\"a et al.\ 2009a).
Similar results had been obtained earlier in the geodynamo context 
(e.g.\ Jones \& Roberts 2000; Rotvig \&
Jones 2002), but subsequent simulations by Cattaneo \& Hughes (2006)
suggested that large-scale field generation becomes impossible at
larger magnetic Reynolds numbers.
However, this is now believed to be a consequence of 
rotation being too slow (K\"apyl\"a et al.\ 2009a).

Furthermore, during the last few years simulations have verified that
shear can be
the sole driver of large-scale dynamo action in non-helical turbulence
(Yousef et al.\ 2008a,b;
Brandenburg et al.\ 2008). The shear-current effect (Rogachevskii \&
Kleeorin 2003, 2004) has been proposed as a mechanism responsible for
the dynamo, but the details of the numerical results disagree
in various respects with the theoretical expectations.
This numerical
finding provided a motivation to add large-scale shear into a
convective system (K\"apyl\"a et al.\ 2008, 2009c; Hughes \& Proctor
2009). 
As a result it was found that these models indeed yield strong
large-scale dynamos, but their theoretical interpretation in terms of
mean-field theory is not as clear as in the case where only rotation
is present (K\"apyl\"a et al.\ 2009b).

Local simulations can be used to demonstrate the existence of a
large-scale dynamo and the processes that are responsible for the
generation of the magnetic fields can be more readily studied. However, they
cannot be used to study the global behaviour of solar and stellar
dynamos, which occur in spherical geometry and generate a much
richer variety of physical effects. Furthermore, the large-scale
flows that need to be imposed in the local models can arise
self-consistently in the spherical models.

Simulations of convective dynamos in spherical geometry have a long
history starting from the seminal papers of Gilman
(1983) and Glatzmaier (1985), to the present 
(e.g.\ Brun et al.\ 2004; Browning et al.\ 2006;
Brown et al.\ 2007, 2009). These models are now quite advanced and,
with suitable boundary conditions, are capable of reproducing the
solar internal rotation profile (e.g.\ Miesch et al.\ 2006). However,
no cyclic large-scale dynamos have so far been discovered
in models of the solar convection zone
(e.g.\ Brun et al.\ 2004; Browning et al.\
2006). On the other hand, models of rapidly rotating stars do exhibit 
significant
large-scale magnetic fields (Brown et al.\ 2007, 2009), although none
of these studies has yet been able to produce poleward propagating
dynamo waves first observed by Gilman (1983), or anything akin 
to the solar cycle.

Modelling a full convection zone in a spherical shell 
though clearly desirable, has an important drawback,
with the spatial resolution achieved in such simulations 
being inevitably lower than in models that encompass only a part of the
shell.
One can also argue that it may not be necessary to model the full
latitudinal and longitudinal extents of the convection zone in order 
to capture the essential physics of the solar dynamo. 
Main arguments to justify this approach are that the solar magnetic
activity appears to be concentrated in a latitude range of
$\pm40\degr$ around the equator and that the large-scale field is
mostly axisymmetric.
This has recently motivated simulations of forced turbulence 
in parts (`wedges') of full spherical shells
(Brandenburg et al.\ 2007; Mitra et al.\ 2009a,b). 
These simulations aim to strike a reasonable compromise 
between the requirements of higher spatial resolution on the one hand
and globality on the other.
Convection simulations in wedge-shaped geometry have already been
performed by Robinson \& Chan (2001) and DeRosa \& Hurlburt (2003).
Those simulations, however, did not include magnetic fields.
In the present paper we report preliminary results of convection-driven 
dynamos in spherical wedge-geometry.

\section{Model}

The simulations reported here are performed in spherical 
polar coordinates where $(r,\theta,\phi)$ denote radius, colatitude, and 
the azimuth, respectively (in some of our space-time diagrams, 
e.g.\ Fig.~\ref{buttwedge1}, we use the latitude $\Theta=90\degr-\theta$.)
As our domain we consider, instead of a full
spherical shell, a spherical wedge that spans $0.6R \le r \le R$ in
radius (where $R$ is the outer radius of the sphere which is also used as
our unit length), $-\Delta \Theta <\Theta < \Delta \Theta$ in latitude,
and $0<\phi<\Delta \phi$ in longitude. 
We then solve the following set of compressible magnetohydrodynamic
equations
for the magnetic vector potential $\bm{A}$ (which is related to
the magnetic field $\bm{B}$ by $\bm{B}=\bm\nabla\times\bm{A}$), 
the logarithmic density
$\ln\rho$, the velocity $\bm{U}$, and the specific entropy $s$:
\begin{equation}
\frac{D \bm A}{Dt} = \bm{U}\times\bm{B} -\mu_0\eta {\bm J}, \label{equ:AA}
 \end{equation}
\begin{equation}
\frac{D \ln \rho}{Dt} = -\DIV{\bm U},
\end{equation}
\begin{equation}
\frac{D \bm U}{Dt} =
{\bm g} -2\bm{\Omega}\times{\bm U} 
+ \frac{1}{\rho}\left(\bm{J}\times\bm{B} \! - \! {\bm \nabla}p
\! + \! \bm{\nabla} \cdot 2 \nu \rho \bm{\mathsf{S}}\right), \label{equ:UU}
\end{equation}
\begin{equation}
T\frac{D s}{Dt} = \frac{1}{\rho} \bm{\nabla} \cdot K \bm{\nabla}T + 2 \nu \bm{\mathsf{S}}^2 + \frac{\mu_0\eta}{\rho}\bm{J}^2, \label{equ:ene}
\end{equation}
where $D/Dt = \pd/\pd t + \bm{U} \cdot \bm{\nabla}$ is the advective
derivative, 
and $\bm{J}=\mu_0^{-1}\bm\nabla\times\bm{B}$ is the current density. The
vacuum permeability is given by $\mu_0$, whereas magnetic diffusivity
and kinematic viscosity are given by $\eta$ and $\nu$, respectively.
The rotation vector is given by
$\bm\Omega=\Omega_0(\cos\theta,-\sin\theta,0)$, where $\Omega_0$ is
the constant rotation rate of the frame.
For simplicity, we have ignored the centrifugal force, as it is a
second order effect in $\Omega$;
see Dobler et al.\ (2006) for a more detailed justification.
The gravitational acceleration, $\bm{g}$, is given by
\begin{equation}
\bm{g} = - \frac{GM}{r^2}\hat{\bm{r}},
\end{equation}
where $G$ is the universal constant of gravitation, $M$ is the mass of the
star, and $\hat{\bm r}$ is a unit vector in the radial direction. 
The fluid obeys an ideal gas law $p=\rho e (\gamma-1)$, where
$p$ and $e$ are the pressure and internal energy, respectively, and
$\gamma=c_{\rm P}/c_{\rm V} = 5/3$ is the ratio of specific heats at
constant pressure and volume. The heat conductivity is given by $K$ and
the internal energy is related to the
temperature via $e=c_{\rm V} T$.
The specific entropy is given by
\begin{eqnarray}
s=c_{\rm V}\ln(p/p_0)-c_{\rm P}\ln(\rho/\rho_0),
\end{eqnarray}
where $p_0$ and $\rho_0$ are constants defined below.
The temperature can now be written as
\begin{eqnarray}
T=T_0 \exp\left[\gamma\frac{s}{c_{\rm P}} + (\gamma-1)\ln\left(\frac{\rho}{\rho_0}\right) \right],
\end{eqnarray}
where $T_0=p_0/[c_{\rm V}\rho_0(\gamma-1)]$.

The rate of strain tensor $\bm{\mathsf{S}}$ is given by
\begin{eqnarray}
\mathsf{S}_{ij} = \onehalf (U_{i;j}+U_{j;i}) - \onethird \delta_{ij} \DIV \bm{U}\;
\end{eqnarray}
where `;' denotes covariant differentiation (see Mitra et al.\ 
2009a, for more details).
We consider a setup where the stratification in the region below $r=0.7R$ is
convectively stable, whereas the region above is unstable. The
positions of the bottom of the domain, bottom of the convectively
unstable layer, and the top of the domain are given by
$(r_1,r_2,r_3)=(0.6,0.7,1)R$. The setup
yields a hydrostatic equilibrium temperature
\begin{equation}
T(r)=T_3+\int_{r_3}^{r}\frac{|\bm{g}|}{c_{\rm V}(\gamma-1)(m+1)}dr,
\end{equation}
where $T_3$ is the temperature at $r_3$ and where $m=m(r)$ is chosen to have
a suitable radial profile. This procedure gives a temperature stratification
for which the logarithmic temperature
gradient (not to be confused with the gradient operator $\bm{\nabla}$)
is given by
\begin{equation}
\nabla \equiv \pd \ln T/\pd \ln p = (m+1)^{-1},
\end{equation}
The stratification is convectively unstable when $\nabla-\nabla_{\rm
  ad}>0$, where $\nabla_{\rm ad}=1-1/\gamma$. Thus $m$ can be
considered as a polytropic index, with values $m<1.5$ yielding an
unstable stratification. We choose $m=1$ in the convectively unstable
layer ($r>r_2$) and $m=3$ in the stable layer below ($r_1<r<r_2$)
with a smooth transition in between. The density stratification is
obtained by requiring the hydrostatic equilibrium condition to be
satisfied.

\begin{figure}
\resizebox{\hsize}{!}
{\includegraphics{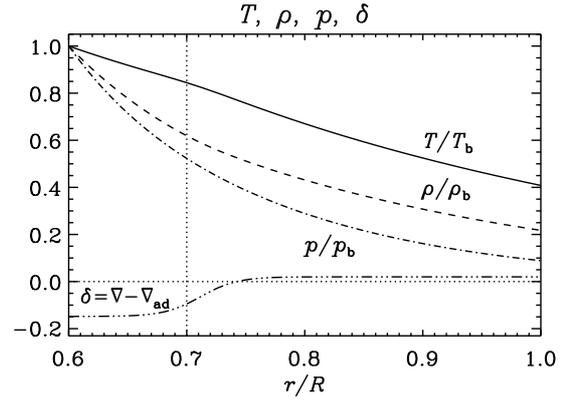}}
\vspace{-.2cm}
\caption{Initial stratification of temperature (solid line), density
  (dashed), pressure (dot-dashed), and the superadiabatic temperature
  gradient (dash-triple-dotted). The subscripts $b$ refer to the values
  at $r=r_1$.}
\label{pstrat}
\end{figure}

The thermal conductivity is obtained by requiring a constant
luminosity $L$ throughout the domain via
\begin{equation}
K=\frac{L}{4\pi r^2 \pd T/\pd r}.
\end{equation}
In order to speed up thermal relaxation, we use $m=1$ only for the
thermal conductivity profile, whereas the actual thermal
stratification of the thermodynamic variables followed a shallower
profile with $\rho\sim T^{1.4}$. The
initial non-convecting stratification is shown in Fig.~\ref{pstrat}.
A fixed temperature gradient is imposed at the base of the domain
which leads to a constant flux of energy into the domain. At the outer
radial boundary the temperature is kept constant and on the
latitudinal boundaries we set $\pd_\theta T=0$ so that there is no
heat flux in or out of the domain.

For the velocity, the upper radial and the latitudinal boundaries are
taken to be impenetrable and stress free, whereas at the lower boundary 
rigid rotation is enforced by taking
\begin{eqnarray}
\lefteqn{U_r= U_\phi =0, \frac{\pd U_\theta}{\pd r}=\frac{U_\theta}{r} \,\,\quad
\quad \quad \quad \quad (r=r_1),}\\
\lefteqn{U_r=0, \frac{\pd U_\theta}{\pd r}=\frac{U_\theta}{r}, \frac{\pd
U_\phi}{\pd r}=\frac{U_\phi}{r} \,\,\, \quad \quad (r=r_3),}\\
\lefteqn{\frac{\pd U_r}{\pd \theta}=U_\theta=0, \frac{\pd U_\phi}{\pd
\theta}=U_\phi \cot \theta \quad \quad (\theta=\theta_1,\theta_2),}
\end{eqnarray}
where $\theta_1=90\degr-\Delta\Theta$ and $\theta_2=90\degr+\Delta\Theta$. 
We use perfect conductor boundary conditions for the magnetic field at
the lower radial and latitudinal boundaries, and a normal-field condition
on the radial top boundary. In terms of the magnetic
vector potential these translate to
\begin{eqnarray}
\lefteqn{\frac{\pd A_r}{\pd r}= A_\theta=A_\phi =0 \,\quad \quad \quad \quad
\quad \quad \quad \quad  (r=r_1),}\\
\lefteqn{A_r=0, \frac{\pd A_{\theta}}{\pd r}=-\frac{A_{\theta}}{r}, \frac{\pd
A_{\phi}}{\pd r}=-\frac{A_{\phi}}{r}\quad\,  (r=r_3),}\\
\lefteqn{A_r=\frac{\pd A_\theta}{\pd\theta}=A_\phi=0 \quad \quad \quad \quad
\quad \quad (\theta=\theta_1,\theta_2).}
\end{eqnarray}
Finally, all quantities are  taken to be periodic in the azimuthal direction.

\begin{figure*}
\resizebox{0.925\hsize}{!}
{\includegraphics{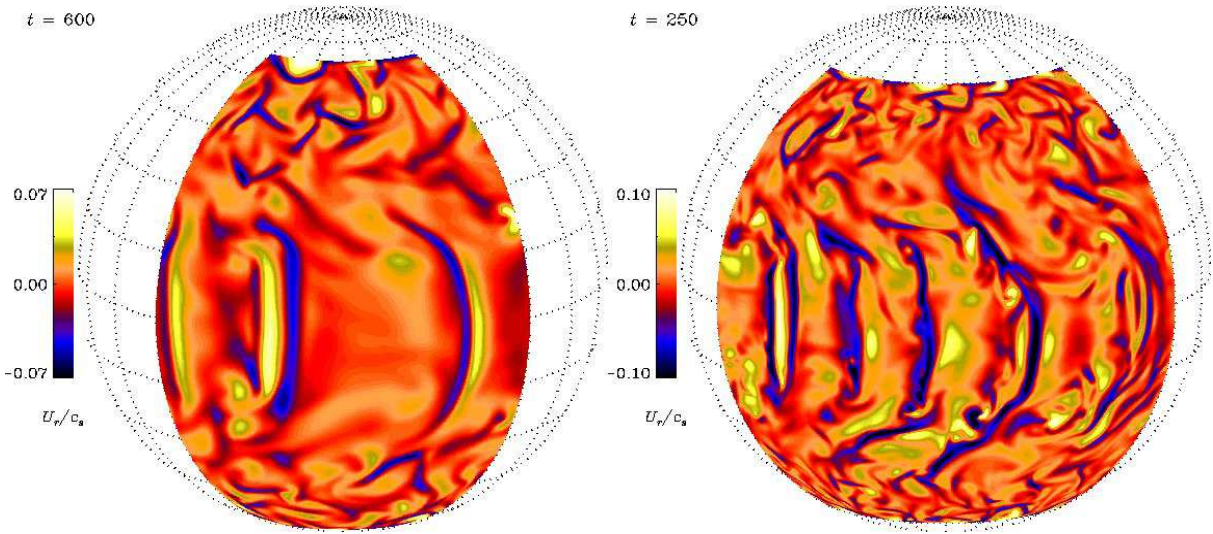}}
\caption{ (online colour at: www.an-journal.org) Radial velocity $U_r$
  near the upper boundary ($r=0.98R$) from Runs~A4 (left) and B1 (right). 
  The slices are
  projected onto a spherical surface which is tilted $15\degr$ towards
  the observer. The dotted lines in the background show a grid in latitude
  and longitude in increments of $15\degr$. See also \texttt{http://www.helsinki.fi/\ensuremath{\sim}kapyla/movies.html}}
\label{slices_ur}
\end{figure*}

The simulations were performed using the {\sc Pencil Code}%
\footnote{\texttt{http://pencil-code.googlecode.com/}},
which uses sixth-order explicit finite differences in space and a third-order
accurate time stepping method (see Mitra et al.\ (2009a) for 
further information regarding the adaptation of the {\sc Pencil Code}
to spherical coordinates).
We use grid sizes of $64\times192\times128$ (Runs~A1--A4)
and $64\times256^2$ (Run~B1) where the angular extent,
$2\Delta\Theta\times\Delta\phi$, is either $2\times67.5\degr\times90\degr$
or $2\times60\degr\times120\degr$; see Table~\ref{runs} below.
Typical snapshots of the radial velocity $U_r$ for the two domain
sizes are shown in Fig.~\ref{slices_ur}.

\subsection{Units, nondimensional quantities, and parameters}
Dimensionless quantities are obtained by setting
\begin{eqnarray}
R = GM = \rho_0 = p_0 = c_{\rm P} = \mu_0 = 1\;,
\end{eqnarray}
where $\rho_0$ is the density at $r_1$. The units of length, time,
velocity, density, entropy, and magnetic field are then given by
\begin{eqnarray}
\lefteqn{ [x] = R\;,\;\; [t] = \sqrt{R^3/GM}\;,\;\; [U]=\sqrt{GM/R}\;,\;\;} \nonumber \\ 
\lefteqn{  [\rho]=\rho_0\;,\;\;[s]=c_{\rm P}\;,\;\;
[B]=\sqrt{\rho_0\mu_0GM/R}\;. }
\end{eqnarray}
The simulations are now governed by the dimensionless Prandtl,
Reynolds, Rayleigh and Coriolis numbers defined by
\begin{eqnarray}
\Pra&=&\frac{\nu}{\chi_0}\;,\;\;\Pm=\frac{\nu}{\eta},\\
\Rey&=&\frac{\urms}{\nu \kef}  \;,\;\; \Rm=\frac{\urms}{\eta \kef}=\Pm\,\Rey,\\
\Ra&=&\frac{GM(r_3-r_2)^4}{\nu \chi_0 R^2} \bigg(-\frac{1}{c_{\rm P}}\frac{{\rm d}s}{{\rm d}r} \bigg)_{r_{\rm m}}\;,\\
\Co&=&\frac{2\,\Omega_0}{\urms \kef}\;,
\end{eqnarray}
where $\chi_0 = K/(\rho_{\rm m} c_{\rm P})$ is the thermal
diffusivity, $\kef=2\pi/(r_3-r_2)$ is an estimate of the wavenumber of
the energy-carrying eddies, $\rho_{\rm m}$ is the density in the
middle of the unstable layer at \ $r_{\rm m}=\onehalf(r_3-r_2)$,
and $\urms$ is the volume averaged rms velocity. The
entropy gradient, measured at $r_{\rm m}$ in the non-convecting
initial state, is given by
\begin{eqnarray}
\bigg(-\frac{1}{c_{\rm P}}\frac{{\rm d}s}{{\rm d}r}\bigg)_{r_{\rm m}} = \frac{\nabla_{\rm m}-\nabla_{\rm ad}}{H_{\rm P}}\;,
\end{eqnarray}
where $\nabla_{\rm m} = (\pd \ln T/\pd \ln p)_{r_{\rm m}}$, 
and $H_{\rm P}$ is the pressure scale height at $r_m$.

The energy that is deposited into the domain from the base is
controlled by the luminosity parameter
\begin{equation}
\mathcal{L} = \frac{L}{\rho_0 (GM)^{3/2} R^{1/2}}.
\end{equation}
Furthermore, the stratification is determined by the pressure scale
height at the surface
\begin{eqnarray}
\xi = \frac{c_{\rm V}(\gamma-1) T_3}{GM/R}.
\end{eqnarray}
Similar parameter definitions were used earlier by Dobler et al.\ (2006).
The parameter values used in our runs are listed in Table~\ref{runs}.

\begin{table}
 \centering
 \caption{Summary of the runs. $\Rm$ and $\Co$ are given in the saturated 
   state of the dynamo. $\Pm=1$ in all runs.}
\label{runs}
\begin{tabular}{lcccccc}\hline
Run & $\Rm$ & $\Co$ & $\Pr$ & $\Ra$ & $\mathcal{L}$ & $\xi$ \\
\hline
A1  & 36    & 0    & 0.2   & $1.2\cdot10^5$ & $10^{-3}$ & 0.084 \\
A2  & 71    & 0.3  & 0.2   & $1.2\cdot10^5$ & $10^{-3}$ & 0.084 \\
A3  & 35    & 1.3  & 0.2   & $1.2\cdot10^5$ & $10^{-3}$ & 0.084 \\
A4  & 28    & 2.5  & 0.2   & $1.2\cdot10^5$ & $10^{-3}$ & 0.084 \\
\hline
B1  & 45    & 5.5  & 0.67  & $2.6\cdot10^6$ & $1.2\cdot10^{-4}$ & 0.078 \\
\hline
\end{tabular}
\end{table}

\begin{figure*}
\resizebox{.925\hsize}{!}
{\includegraphics{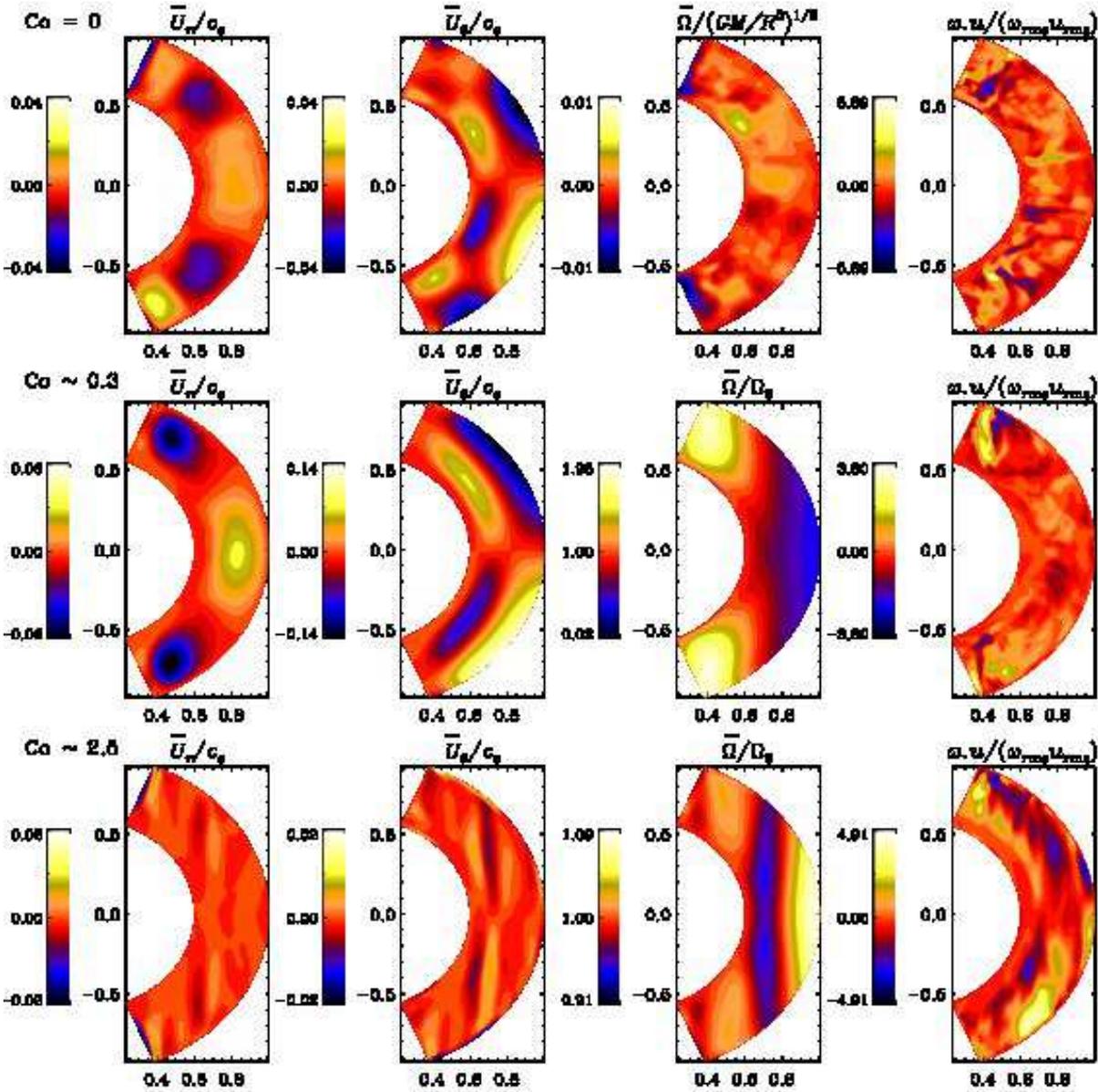}}
\vspace{-.15cm}
\caption{(online colour at: www.an-journal.org) From left to right: azimuthally averaged $\mean{U}_r$,
  $\mean{U}_\theta$, $\mean{\Omega}=\mean{U}_\phi/(r \sin \theta)+\Omega_0$,
  and kinetic helicity $\mean{\bm\omega\cdot\bm{u}}$ from runs with
  $\Co=0$ (top row, Run~A1), $\Co\approx0.3$ (middle, Run~A2) and $\Co\approx2.5$
  (bottom, Run~A4) from the kinematic stages of the simulations.
  Here $\omega_{\rm rms}$ is the rms-value of the vorticity, 
  $\bm\omega=\CURL\bm{u}$.}
\label{puu}
\end{figure*}

\section{Results}
\label{sec:results}

\subsection{Hydrodynamic solutions}

Before discussing our results with magnetic fields
we briefly describe a limited range of
hydrodynamic solutions. One of the most important reasons for using
spherical geometry is that large-scale shear and meridional flows can
be generated self-consistently when the overall rotation of the star is
included. In such a setting the kinetic helicity is also likely to vary
as a function of position.
At large magnetic Reynolds numbers,
this information can then be used as a proxy
of the trace of the $\alpha$ tensor of mean-field dynamo theory 
(e.g.\ R\"adler 1980).
In order to study
the effects of rotation on the flow, we study here a range of models where the
rotation rate is gradually increased.

Figure~\ref{puu} shows cross-sections of longitudinal and temporal
averages of the meridional velocities, rotation profile, and
kinetic helicity for the kinematic stages of the simulations, 
from three runs with varying degrees of rotational influence. In
the absence of rotation (Run~A1), the meridional velocity components show a
large-scale structure which is related to long-lived convective
cells. However, as there is no mean rotation, no differential rotation 
is generated in this case,
and, although the kinetic helicity $\mean{\bm\omega\cdot\bm{u}}$
shows fairly large peak values, there is no systematic component
present. These results are in accordance with theoretical
considerations for nonrotating convection.

When rotation is relatively weak, with $\Co\approx0.3$ (Run~A2), very strong
differential rotation and meridional flow is generated. The
resulting differential rotation is anti-solar, i.e.\ the equator is rotating
slower than the high-latitude regions. The difference between
minimum and maximum rotation rates is remarkably large -- about twice
the rotation rate of the star. The meridional circulation is also very
strong and oriented anti-clockwise. A similar result has been reported
by Robinson \& Chan (2001) for their convection simulations in a
wedge-shaped domain. However, in their case the rotation profile
changed to a more solar-like configuration when the
longitudinal extent of the domain was doubled.
We have not experimented with
increasing the $\phi$-extent, but we note that there are
some other differences; for example, our domain is
wider in longitude than that of Robinson \& Chan (2001). The kinetic
helicity in this case shows signs of large-scale structure: in the northern
hemisphere there appears to be positive helicity near the equator and
negative helicity at higher latitudes. A similar profile, but with
opposite sign, is discernible in the southern hemisphere. The
strongest features appear near the latitudinal boundaries and are
possibly due to the converging large-scale flows that occur there.

In the rapidly rotating case with $\Co\approx2.5$ (Run~A4) 
the relative differential rotation is significantly
reduced and equatorial acceleration is observed.
However, the absolute value of ${\rm d}\Omega/{\rm d}r$, relevant for the
large-scale dynamo, is of the same order of magnitude, or somewhat
larger near the equator, than in Run~A2 with $\Co\approx0.3$.
In agreement with earlier simulations of convection in rotating
spherical shells (e.g., Miesch et al.\ 2000) the rapid rotation regime
is dominated by large prograde `banana cells' in the radial velocity;
see Fig.~\ref{slices_ur}.
In our most rapidly rotating models we also see nonaxisymmetric
structures in the equatorial regions.
Brown et al.\ (2008) described such structures as `nests' of convection.
The strength of convection in those
structures varies periodically in time and the maxima coincide with enhanced
differential rotation similarly as reported by Busse (2002).
The left panel of Fig.~\ref{slices_ur} shows indications of localised
convection whereas in the higher Rayleigh number model (Run~B1, right
panel) these features have not yet developed. We also find that these
localised convection structures are suppressed when magnetic fields of
equipartition strength are present.
The contours of constant $\Omega$ are aligned
with the rotation vector, in
accordance with the Taylor--Proudman balance
\begin{equation}
\frac{\pd\Omega^2}{\pd z}\approx 0,
\end{equation}
where the $z$ derivative is evaluated as
\begin{equation}
\frac{\pd}{\pd z}=\cos\theta \frac{\pd}{\pd r} -
\frac{\sin \theta}{r}\frac{\pd}{\pd \theta}.
\end{equation}
Only weak meridional circulation is observed and the flow consists of
several small cells. This can also be attributed to the
Taylor-Proudman balance and has also been found in mean-field
calculations (Brandenburg et al.\ 1991).
In the present models the Taylor--Proudman
balance always dominates the rotation profile in the rapid rotation
regime. 
Preliminary results from a run with roughly 20 times higher Rayleigh
number (Run~B1) with $\Co\approx5.5$ gives qualitatively very similar
results.
Reproducing the solar rotation profile self-consistently
without resorting to an imposed temperature difference between the
equator and the pole (e.g.\ Miesch et al.\ 2006) or including
unresolved turbulent convective heat fluxes (e.g.\ Durney \& Roxburgh
1971; Kitchatinov et al.\ 1994; R\"udiger et al.\ 2005) is likely to
be difficult.

We enforce rigid rotation at the lower boundary at $r=0.6\,R$, although this
is unphysical and not even enough to avoid spreading the differential rotation
into the stably stratified layer below the convection zone 
between $r=0.6\,R$ and $r=0.7\,R$ (see also Miesch et al.\ 2000). This
spreading is due to the viscosity which is much larger in our
simulations than in the solar convection zone. Thus, no tachocline
forms in our models. It is also likely that a self-consistent formation
of a tachocline requires such high numerical resolution that the
problem cannot be tackled successfully in the foreseeable future. A
more practical solution is probably the damping of the velocities in the
overshoot region, similarly as in Browning et al.\ (2006).

We also find that in the most rapidly rotating case considered here,
the kinetic helicity has a well-defined profile with a mostly negative
(positive) value in the northern (southern) hemisphere in the outer
layers of the convectively unstable layer. Furthermore, there is a
sign change as a function of radius near the base of the convection
zone. These results are in accordance with theoretical considerations
and earlier results from local and global simulations \ (e.g.\ \
Miesch et al. 2000; Brun et al. 2004; K\"apyl\"a et al. 2006). 
The simple first-order smoothing 
relation that is valid for isotropic turbulence states that 
$\alpha\propto-\mean{\bm{\omega}\cdot\bm{u}}$ which suggests that 
the $\alpha$ effect should in the present case have a positive 
(negative) value in the northern (southern) hemisphere.
This will be examined in more detail in the next section.

\begin{figure}
\resizebox{.925\hsize}{!}
{\includegraphics{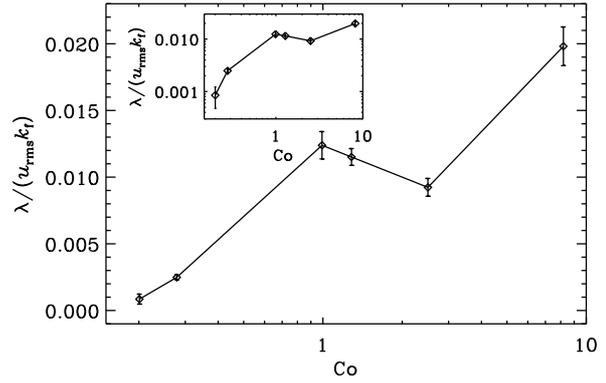}}
\vspace{-.2cm}
\caption{Growth rate of the magnetic field as a function of rotation.}
\label{pgrw}
\end{figure}

\subsection{Dynamo solutions}
We begin our study of dynamo excitation with 
a model with $\Rm\approx36$ that does not act as a dynamo in 
the absence of rotation ($\Co=0$).
For $\Co\approx0.2$ the
magnetic field shows exponential growth. The time averaged growth rate
of the magnetic field,
\begin{equation}
\lambda = \langle {\rm d} \ln \brms/{\rm d}t\rangle_t,
\end{equation}
increases with rotation for small values of
$\Co$, see Fig.~\ref{pgrw}, although it shows a dip at $\Co\approx2.5$.

\begin{figure}
\resizebox{.925\hsize}{!}
{\includegraphics{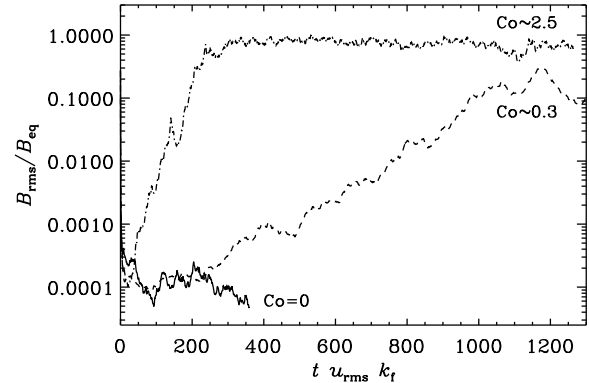}}
\vspace{-.15cm}
\caption{Root mean square value of the magnetic field as a function of
  time for runs with $\Co=0$ (solid line, Run~A1), $\Co\approx0.3$ (dashed, Run~A2),
  and $\Co\approx2.5$ (dash-dotted, Run~A4).}
\label{pbrms}
\end{figure}

\begin{figure}
\resizebox{.925\hsize}{!}
{\includegraphics{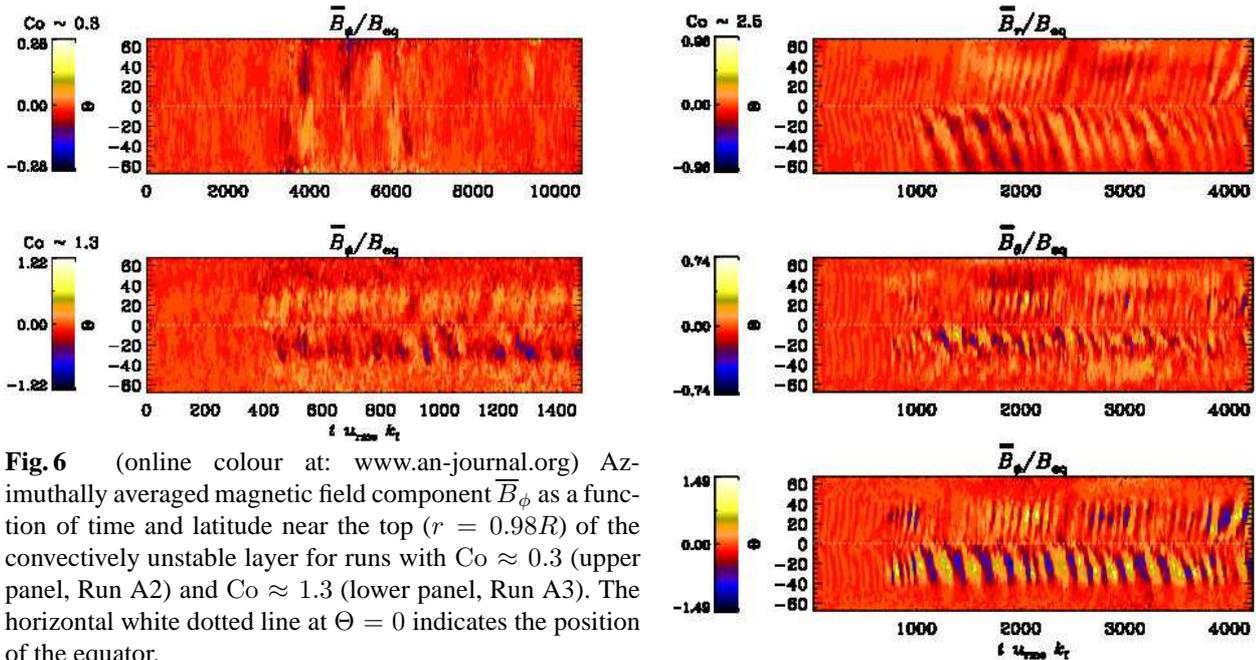}}
\vspace{-.15cm}
\caption{(online colour at: www.an-journal.org) Azimuthally averaged magnetic field component $\mean{B}_\phi$
  as a function of time and latitude near the top 
   ($r=0.98R$) of the 
  convectively unstable layer for runs with $\Co\approx0.3$ (upper 
  panel, Run~A2) and $\Co\approx1.3$ (lower panel, Run~A3).
  The horizontal white dotted line at $\Theta=0$ indicates the 
  position of the equator.
}
\label{pbfly}
\end{figure}

In the slowly rotating cases ($\Co\la0.3$) the saturation level of the
magnetic field is rather low, $\brms/\Beq\approx0.1$, where
$\Beq=\sqrt{\mu_0\rho\urms^2}$ is the equipartition field
strength (see Fig.~\ref{pbrms}). 
Furthermore, there is a large-scale axisymmetric component to the field
which is thus well described by a longitudinal average. In Run~A2
with $\Co\approx0.3$ the large-scale field is approximately
antisymmetric with respect to the equator and shows irregular sign changes on a
long time scale (see the upper panel of
Fig.~\ref{pbfly}).

\begin{figure}
\resizebox{.925\hsize}{!}
{\includegraphics{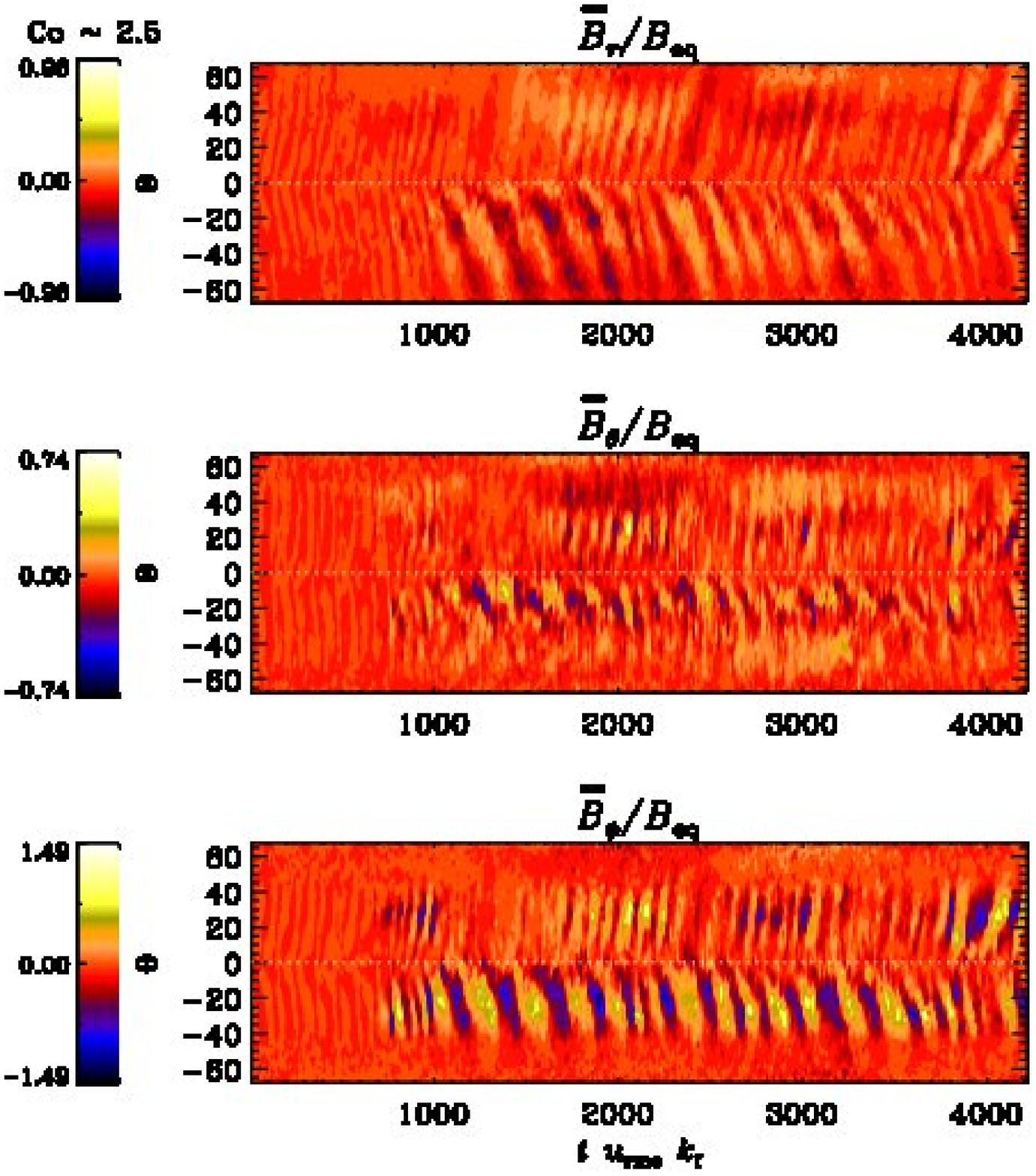}}
\vspace{-.15cm}
\caption{(online colour at: www.an-journal.org) Magnetic field components $\mean{B}_r$ (top panel),
  $\mean{B}_\theta$ (middle), and $\mean{B}_\phi$ (bottom) averaged
  over the azimuthal direction as a function of time and latitude near the top
  ($r=0.98R$) of the convectively unstable layer from Run~A4.
}
\label{buttwedge1}
\end{figure}

\begin{figure*}
\resizebox{.925\hsize}{!}
{\includegraphics{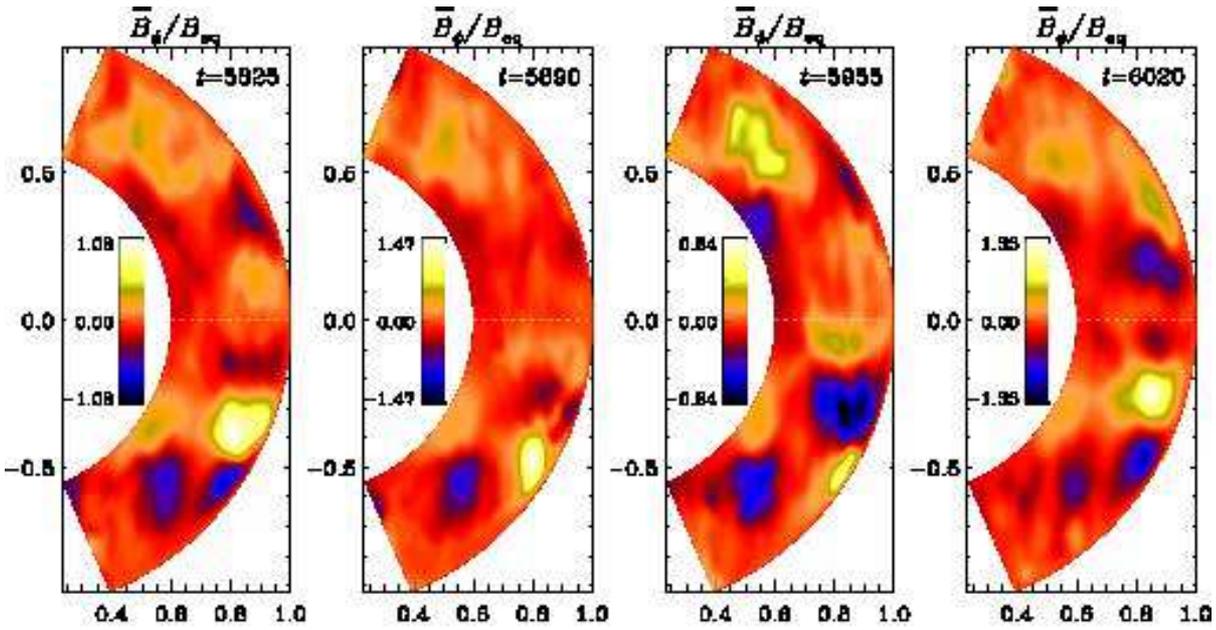}}
\vspace{-.7cm}
\caption{(online colour at: www.an-journal.org) Azimuthally averaged magnetic field component $\mean{B}_\phi$
  from four times separated by $\Delta t=65\sqrt{R^3/GM}$,
  corresponding to $\Delta t \urms \kef\approx41$, from Run~A4 with
  $\Co\approx2.5$.}
\label{pslices}
\end{figure*}

\begin{figure}
\resizebox{.925\hsize}{!}
{\includegraphics{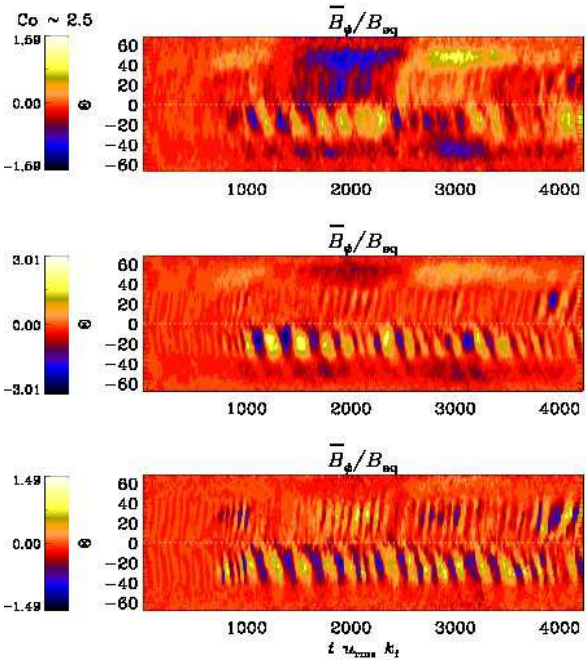}}
\vspace{-.15cm}
\caption{(online colour at: www.an-journal.org) Magnetic field component $\mean{B}_\phi$, averaged over the
  azimuthal direction as a function of time and latitude from the bottom (top
  panel, $r=0.7R$), middle (middle panel,$r=0.85R$) and near the top
  (bottom panel, $r=0.98R$) of the convectively unstable layer from Run~A4.
}
\label{buttwedge2}
\end{figure}

As the rotation rate is increased the qualitative character
of the differential rotation changes from anti-solar to solar-like,
as was seen in the previous
section. A qualitative change occurs also in the behaviour of the
dynamo so that for $\Co\approx1.3$ (Run~A3) magnetic fields of the order of the
equipartition strength are obtained.
Furthermore, the fields show a
large-scale pattern that stays roughly independent of time, i.e.\ no
sign changes are found. However, near the surface there are
indications of a time-variable large-scale structure (lower panel of
Fig.~\ref{pbfly}).
When the
rotation rate is doubled ($\Co\approx2.5$, Run~A4), the magnetic field indeed
shows oscillatory behaviour near the equator, see
Figs.~\ref{buttwedge1} and \ref{buttwedge2}, with branches of
magnetic fields appearing to move from the equator towards higher
latitudes. We interpret the phenomenon as a dynamo wave propagating
towards the poles. More precisely, assuming that the sign of the
$\alpha$ effect is opposite to that of the kinetic helicity, we would thus
expect $\alpha$ to have a positive (negative) value in the northern 
(southern) hemisphere. 
Furthermore,
there is a positive ${\rm d}\Omega/{\rm d}r$ near the equator, which leads to
$\alpha\,{\rm d}\Omega/{\rm d}r>0$ which, based on mean-field theory, should lead to
poleward migration (e.g.\ Yoshimura 1975).
We note that similar results were reported by Gilman (1983).
However, a more detailed interpretation of the origin of the dynamo
requires a better understanding of the turbulent transport coefficients.

\begin{figure*}
\resizebox{0.925\hsize}{!}
{\includegraphics{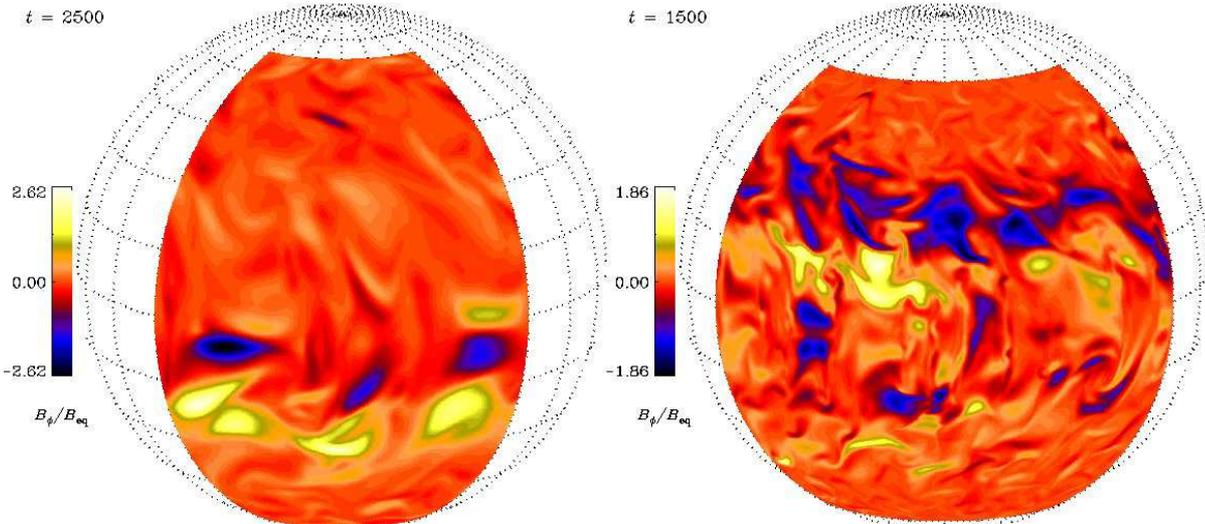}}
\vspace{-.3cm}
\caption{ (online colour at: www.an-journal.org) Toroidal magnetic
  field $B_\phi$ near the upper boundary ($r=0.98R$) from the Runs~A4 (left)
  and B1 (right).}
\label{slices_bt}
\end{figure*}

We note that in the kinematic stage of the simulation the period of
the oscillations is well-defined and the sign changes occur nearly
similarly on both hemispheres. In the saturated stage the oscillation
period is expected to become slightly longer when the implied turbulent diffusion is
quenched by the magnetic field (e.g.\ K\"apyl\"a \& Brandenburg
2009). Such behaviour is indeed observed, but it appears that the two
hemispheres begin to show a distinct asymmetry with different oscillation
periods and magnitudes of the fields in the saturated state (see
Figs.~\ref{buttwedge1}, \ref{pslices}, and \ref{slices_bt}). 
We note that hemispherical dynamos, where one of the hemispheres is
almost devoid of magnetic fields, have been reported from rapidly
rotating spherical shells (Grote \& Busse 2000; Busse 2002). These
dynamos also exhibit poleward migration of activity belts similar to
the results reported here.
Furthermore, Fig.~\ref{buttwedge2} shows
that in the deeper layers the magnetic field has a very different
structure with strong large-scale fields reaching much higher
latitudes that can also change sign, but more irregularly and on a time
scale that is much longer than that shown by the field near the
surface. Since the differential rotation in these models is
appreciable only near the equator, it is conceivable that there is a
competition between different dynamo modes in the deeper layers and
those at the higher latitudes and that the irregular nature of the 
oscillations in the saturated state is caused by the interaction between
these regions.

\section{Conclusions}
\label{sec:conclusions}

In this paper we have presented preliminary results of rotating turbulent convection
in spherical wedge geometry for different rotation rates.
We find that for
small values of the Coriolis number the differential rotation is 
anti-solar, with the equator rotating slower in comparison 
to higher latitudes. When the Coriolis
number is increased above unity, the rotation profile changes to a
more solar-like configuration with faster rotation at the
equator and slower rotation
at higher latitudes. However, with more rapid rotation ($\Co\ga1$) the
Taylor-Proudman balance dictates the rotation profile with contours
that tend to be constant along cylinders.

The same setup is used to study the corresponding dynamo solutions.
The magnetic Reynolds number is chosen such that 
there is no dynamo in the absence of rotation.
A dynamo is excited for sufficiently rapid rotation.
Near the marginally excited state the magnetic
energy is only roughly ten per cent of the kinetic energy,
but a large-scale structure can already be observed. With more rapid rotation
($\Co\ga1$), strong large-scale magnetic fields are obtained. In
this case the field is non-oscillatory.
Increasing the rotation rate further changes the solution such that an
oscillatory magnetic field is observed near the equator in the upper
parts of the convectively unstable layer. The magnetic features
appear at low latitudes and propagate poleward, i.e.\ opposite to the
behaviour seen in the Sun. However, the kinetic helicity in this run
suggests that the $\alpha$ effect is positive (negative) in the
northern (southern) hemisphere. This together with the positive radial
gradient of $\Omega$ near the equator is consistent with a dynamo wave
propagating poleward.

Our model is in some ways similar to that of
Gilman (1983) which also omitted the regions near the pole and used the
same boundary conditions for the magnetic field. The main difference
between our models is that we include density stratification and add a
lower overshoot layer.
Many of the results, e.g.\ a poleward propagating dynamo wave in our
most rapidly rotating runs, are qualitatively similar to those found by
Gilman (1983). However, we obtain a deep-seated non-oscillatory field
and see large differences in the activity levels on different
hemispheres which were not reported by Gilman. The differences between
the setups could explain some of the discrepancies of the results.

Our ultimate aim with this model is to study
dynamo processes that operate in solar and stellar interiors.
In order to achieve this goal we need to find a way to break the
Taylor-Proudman balance. A possible solution would involve the inclusion of
subgrid scale
anisotropic turbulent heat transport in the model. 
It is also unclear how the introduction of latitudinal boundaries 
affects the large-scale behaviour of the solutions.
Issues related to
these problems will be dealt with in future publications.

\acknowledgements{
We thank an anonymous referee for critical remarks that have led
to several improvements of the paper.
The simulations were performed with the computers hosted by CSC, the
Finnish IT center for science financed by the Ministry of Education,
and with computers hosted by QMUL HPC facilities purchased under the
SRIF initiative.
Financial support from the Academy of Finland grants No.\ 121431 (PJK)
and 112020 (MJK) is acknowledged.
This work was supported in part by
the European Research Council under the AstroDyn Research Project 227952
and the Swedish Research Council grant 621-2007-4064.
DM is supported by the Leverhulme Trust.
}

\end{document}